\documentclass[prd,preprintnumbers,nofootinbib,aps,twocolumn]{revtex4-1}
\usepackage{epsfig,amsmath,amssymb}
\usepackage{mathrsfs}
\usepackage[normalem]{ulem}
\usepackage[usenames,dvipsnames]{color}
\usepackage[pagebackref=false, colorlinks=false]{hyperref}

\begin{document}
\title{Holographic Bound on Area of Compact Binary Merger Remnant}

\author{Parthasarathi Majumdar}
\email{bhpartha@gmail.com}
\affiliation{School of Physical Sciences, Indian Association for the Cultivation of Science, Kolkata 700032, India.} 

\author{Anarya Ray}
\email{ronanarya9988@gmail.com}
\affiliation{Department of Physics, University of Wisconsin-Milwaukee, Milwaukee, WI 53201, USA}

\begin{abstract}

Using concomitantly the Generalized Second Law of black hole thermodynamics and the holographic Bekenstein entropy bound embellished by Loop Quantum Gravity corrections to quantum black hole entropy, we show that the boundary cross-sectional area of the post-merger remnant formed from the compact binary merger in gravitational wave detection experiments like GW150914 {\it et. seq.}, by the LIGO-VIRGO collaboration, is bounded from below. This lower bound is more general than the bound obtained from application of Hawking's classical area theorem for black holes, since it does not depend on whether the inspiralling compact binary pair or the postmerger remnant consists of black holes or other exotic compact objects. The derivation of the bound entails an estimate of the entropy of the gravitational waves emitted during the binary merger which adapts to gravitational waves an extant formalism proposed originally for particle ensembles. The results for the minimal cross-sectional area of the merger remnant due to binary compact mergers observed recently by the LIGO-VIRGO collaboration are discussed. While accurate measurement of the mass of the remnant for the BNS merger GW170817 remains a challenge, we provide a {\it proof of principle} that for BNS mergers our lower bound on the cross-sectional area of the remnant provides an alternative approach to probe the validity of neutron star Equations of State, {\it independent} of the measurements of the tidal deformabilities of the components. 
\end{abstract}

\maketitle

\section{Introduction}

Gravitational wave signals have so far been observed by the LIGO and the LIGO-VIRGO collaborations, from inspiralling coalescence of compact binaries \cite{LIGO1}-\cite{LIGO5}, with a  recent report of a possible black hole-neutron star merger \cite{LV18}-\cite{LV19}. A consensus view regarding the earliest observation - GW150914 - of gravitational wave radiation is that it is a consequence of a coalescence of an inspiralling {\it black hole} binary system, with the black holes being Kerr black holes with their mass in the $30 M_{\odot}$ range, and Kerr parameter in the $0.6-0.8$ range. There is debate in the literature that the inspiralling binary system as well as the post-merger remnant may not consist of black holes, since accretion disc observations of xray emission from binary black hole systems with normal stars have never revealed any stellar black hole as massive as the ones reported in GW150914. Boson stars, gravastars and wormholes have been cited as possible alternatives, categorized collectively as Exotic Compact Objects \cite{giu16}-\cite{cardos17}. If such compact gravitating objects accrete material from stellar matter and interstellar dust in their vicinity, then, it has been argued in ref.\cite{yunes19}, using Thorne's Hoop conjecture \cite{thorne}, that such exotic compact configurations become gravitationally unstable, collapsing to a black hole. This conclusion apparently depends on certain assumed details of the accretion process.

Now, for GW150914, it is not known if the inspiralling binary system actually accretes at all, so that the gravitational instability argument cited above, even if eminently reasonable, may not apply immediately. Nevertheless, if we do assume that the inspiralling system consists of binary black holes, then their merger to a black hole is of course subject to Hawking's Area Theorem: the sum of the horizon areas of the inspiralling black holes must be less than the horizon area of the post-merger black hole remnant. The prediction of the theorem has been shown \cite{pdg}, \cite{badri} to be borne out by the published data on GW150914. However, this classical general relativistic law has been superceded by Bekenstein's Generalized Second Law \cite{bek73} for a universe with black holes, namely that the entropy of the remnant and that of the gravitational waves radiated by the inspiralling binary, must together exceed the sum of the entropies of the two merging black holes. This law is not restricted to classical general relativity, and  is valid within any quantum gravity framework which permits one to compute {\it ab initio} the entropy of macroscopic black holes. The question that comes to mind is : does the Generalized Second Law make a stronger statement on the horizon area of the remnant than the assertion of Hawking's theorem ? The answer is in the affirmative for a quantum gravity framework like Loop Quantum Gravity \cite{ashlew} which indeed permits an ab initio computation of the entropy of isolated black holes \cite{ashbhe97}, \cite{km98}, \cite{pm98}. Such a computation has been shown to yield not only the Bekenstein-Hawking area law for  black hole entropy, but also an entire slew of quantum geometry corrections starting with a term logarithmic in the Bekenstein-Hawking entropy \cite{km2000} - \cite{abhi-pm14} whose coefficient has been argued to be of a `universal' nature \cite{kaul12}. Even though this correction is small, its effect on the minimal cross-sectional area of the remnant is of interest in this case.  

A somewhat more general situation ensues, if the post-merger remnant of the inspiralling black hole pair is not necessarily a black hole. If no assumption is made on the precise astrophysical nature  of the remnant, with simply the information that it must be a compact astrophysical object, we demonstrate in this paper that a constraint may still be derived on the cross-sectional area of the boundary of the remnant. This constraint follows from the Bekenstein entropy bound \cite{bekbd74} embellished or tightened by quantum spacetime corrections alluded to above \cite{dkm2001}. According to the original version of the bound, the entropy of the compact remnant is bounded from above by the entropy of a black hole whose horizon area is identical with the {\it boundary} cross-sectional area of the  remnant. This bound, in its turn, has been established by Bekenstein \cite{bekbd74} on the basis of arguments somewhat analogous to those used by Thorne to establish the Hoop conjecture \cite{thorne}. The basic idea is that a black hole carries the maximum entropy for all {\it compact} astrophysical objects of the same cross-sectional area. Any compact object which does not satisfy the bound to begin with, can adiabatically accrete material from its environment without changing its area, so as to increase its energy/mass to the point where its size starts to fall below its Schwarzschild radius, causing the compact object to be on the verge of gravitational collapse to a black hole. The entropy at this point has also risen adiabatically to the entropy of a black hole, being only a function of its {\it horizon} area which is given by the cross-sectional area of the compact object. Thus, so long as a compact gravitational object has the ability to accrete material from its environment, its entropy is bounded from above by the entropy of a black hole determined by the cross-sectional area of the object. This boundary area of the star prospectively turns into the horizon area of the black hole, once collapse actually occurs. The validity of the bound clearly does not need the actual {\it occurrence} of accretion leading to gravitational collapse to a black hole as a prerequisite. It is adequate that there is a strong possibility of adiabatic accretion of material of the compact star from its environment, without a significant increase in its size.      

Going beyond Bekenstein's hypothesis of the entropy bound, LQG corrections to black hole entropy, inherently holographic in character, provide a {\it tightening} of the entropy bound \cite{dkm2001} for any merger remnant of binary compact coalescence. Thus, even though the entropy bound is hypothetical, albeit one that follows from very general assumptions, its strengthening due to LQG has more solid foundations, where the holographic aspect is embedded within the formalism. The very fact that LQG corrections lead to a stronger entropy bound, rather than invalidating or {\it weakening} it, implies that this bound would very likely result from the full-fledged quantum gravity theory which {\it formally completes} LQG in the future. Because, if the full quantum gravity theory invalidates the entropy bound, then it would most likely contradict basic tenets of LQG. These tenets are well founded and lead, via a systematic analytic procedure, to an {\it ab initio} computation of the quantum-corrected black hole entropy fully consistent with the Generalized Second Law \cite{km2000} - \cite{abhi-pm14}. To reiterate, the hypothetical nature of the bound may invoke questions about the reliability of predictions made from using it. However, the very fact that LQG corrections endow the bound with added precision, should partly allay suspicions about its applicability in principle. The final arbiter on the correctness or otherwise of the quantum-improved bound is of course observational data.      

Our derivation of the bound on the area of the compact remnant follows from concomitant application of the generalized second law and the quantum-corrected entropy bound. As such this will entail an estimate of the entropy of the radiated gravitational waves. This estimate is made based on the assumption that the detected gravitational waves do not scatter substantively enoute to LIGO from the merger source. 

The lower bound on the cross-sectional area of the remnant in any binary compact coalescence has an interesting implication in case of BNS mergers, namely constraining neutron star Equations of State (EoS). The standard approach to this problem entails the measurement of tidal deformations of the component neutron stars in addition to their mass, so that every EoS then relates the mass to the cross-sectional {\it size or radius} of the neutron star. By providing a minimal cross-sectional area of the remnant in terms of the areas of the components, our formulae permits a direct use of every proposed EoS to relate the component masses to their radii. Comparison of our minimal remnant area with the observed remnant area in a BNS merger could then probe the EoS employed. Data on tidal deformations of the neutron star, which is always nontrivial to garner from observations, is not needed in this approach.      

The paper is organized as follows : in the next section the LQG analysis of the entropy of {\it isolated horizons} - as non-stationary generalizations of black holes - is briefly reviewed and the appearance of LQG corrections to the Bekenstein-Hawking area law pointed out. This is followed by a short recap of the holographic nature of LQG corrections  which strengthen the entropy bound. We present our first results on the minimal area of the merger remnant of black hole coalescence in section 3, where the Generalized Second Law and the LQG corrected entropy bound are used simultaneously. The fact that black hole entropy caps the maximum possible entropy of compact stars with areas equal to the horizon area of the black hole, is then used to generalize to the situation where the binary coalescence is not  just of black holes but {\it any} pair of compact objects, including Exotic Compact Objects. This leads to our main result in section 4 for the minimal area of the merger remnant in compact binary coalescence in general, irrespective of their astrophysical nature. This is followed by a section (5) on the estimate of the entropy of gravitational waves radiated by the binary inspiral on their way to merger. The next section (section 6) uses data, i.e., the observed massses of the components of detected BNS merger events from LIGO-VIRGO observations to probe the validity of our lower bound on the the remnant cross-sectional area. Binary black hole (BBH) merger and binary neutron star (BNS) merger are taken as illustrative examples where the theoretical minimum area is cross-checked against the {\it observed} area of the remnant whereever permitted by the data. In the particular case of BNS merger, the existence of a minimal area of the remnant black hole leads to possible upper bounds on the mass of the coalescing neutron stars. This maximum mass, in its turn, further constrains possible Equations of State of neutron stars. As already noted above, this method obviates the necessity to measure tidal deformabilities of coalescing neutron stars, to lead to the EoS. We end in section 7 with a few concluding remarks. 

\section{Quantum spacetime corrections to Bekenstein-Hawking entropy} 

These corrections have been computed within the Loop Quantum Gravity (LQG) formulation of {\it quantum isolated horizons} \cite{ashbhe97} as generalization of event horizons. In this formulation, the classical isolated horizon is treated as a null {\it inner} boundary defined by boundary conditions consistent with the Einstein equation. A Hamiltonian formulation of general relativity with such isolated horizons yields, when the local Lorentz boosts are gauge fixed, a {\it boundary} symplectic structure on the isolated horizon which coincides with that of an $SU(2)$  Chern-Simons theory, of the connection degrees of freedom on the horizon. Now, in the LQG formulation of bulk general relativity, the kinematical description on a spatial slice is given in terms of a spin network \cite{ashlew} whose edges are the holonomies of the $SU(2)$ connection on the slice,  and intertwinners at the vertices are represented by invariant $SU(2)$ tensors. When the bulk spacetime has an isolated horizon as an inner boundary, the edges of the bulk carrying spin {\it puncture} the horizon, depositing their spin at those punctures. So the Chern-Simons degrees of freedom on the horizon interact with these bulk spins as $SU(2)$ point charges. The (kinematical) LQG description of the horizon is then just this : an SU(2) Chern-Simons theory coupled to pointlike $SU(2)$ charges as sources. The Chern-Simons coupling constant $k = A/A_{Pl}$, where $A_{Pl}$ is the Planck area. Thus, large, macroscopic horizon areas correspond to weak coupling of the Chern-Simons theory, permitting an {\it exact} counting of the size of the Hilbert space of the theory in this limit \cite{km98}. This is necessary for an ab initio computation of the microcanonical entropy of the quantum isolated horizon.

The dimensionality of the Hilbert space of the Chern-Simons theory coupled to spins at punctures is itself related to the {\it number} of conformal blocks of the conformally invariant $SU(2)_k$ Wess-Zumino-Witten model that exists on a spatial foliation of the isolated horizon with punctures at the location of the sources. For large $k$, this  number can be computed in terms of the spins \cite{km98} and gives the result, for a spin configuration $j_1, ...j_P$
\begin{eqnarray}
{\cal N}(j_1, ...j_P)  &=& \prod_{i=1}^{P} \sum_{m_i=-j_i}^{j_i} [ \delta_{\sum_{n=1}^P m_n,0} \nonumber \\
&-& \frac12  \delta_{\sum_{n=1}^P m_n,-1} -\frac12 \delta_{\sum_{n=1}^P m_n,1} ] .\label{dimcs}
\end{eqnarray}
The total number of states is given by 
\begin{eqnarray}
{\cal N} = \sum_P \prod_{i=1}^P \sum_{j_i} {\cal N}(j_1, ...j_P).
\end{eqnarray}
Using the standard formula for Boltzmann entropy $S= \log {\cal N}$, in the limit of large $k=A/A_{Pl}$ the following result is obtained for the microcaonical entropy of quantum isolated horizons\cite{km2000}-\cite{abhi-pm14}
\begin{eqnarray}
S_{bh} = S_{BH} - \frac32 \log S_{BH} +{\cal O}(S_{BH}^{-1}) ~, \label{qent}
\end{eqnarray}     
where, $S_{BH} \equiv A/4A_{Pl}$ is the semiclassical Bekenstein-Hawking area law. The holographic nature of the result depicted in (\ref{qent}) is quite apparent from the fact that the entropy of the isolated horizon which generalizes the event horizon of a four dimensional black hole, is eventually computed from the number of conformal blocks of the two dimensional conformal field theory `living' on the 2-sphere obtained as a spatial foliation of the isolated horizon. 

\section{Holographic Entropy Bound} 

\subsection{Generalized Second Law}

According to the Generalized Second Law, if two black holes bh1 and bh2, with horizon cross-sectional areas $A_1$ and $A_2$ respectively,  merge into a compact object ECO with emission of gravitational waves (GW), the entropies of these configurations must obey the inequality
\begin{eqnarray}
S_{ECO} + S_{GW} > S_{bh1}(A_1) + S_{bh2}(A_2) \label{g2l}
\end{eqnarray}
The entropies of the two black holes obey the augmented version of the Bekenstein-Hawking area law, with augmentation due to corrections derived from quantum spacetime fluctuations. 

\subsection{Binary Black Hole Coalescence}

As already mentioned, the original Bekenstein entropy bound was founded by analogy with the Hoop Conjecture. Since addition of energy invariably increases the entropy from a microcanonical standpoint for an {\it isolated} horizon, the entropy of a compact star is bounded from above by the entropy of a black hole whose horizon area coincides with the area of the boundary of the compact star. In other words, $S_{ECO} < S_{bh}(A_E)$, where $A_E$ is the area of the boundary  of the Exotic Compact Object. Since the entropy of a black hole (isolated horizon) is now not just the Bekenstein-Hawking area expression, but the full quantum-corrected entropy (\ref{qent}) for large macroscopic areas, computed holographically as explained earlier, we obtain the holographic Bekenstein bound \cite{dkm2001}
\begin{eqnarray}
S_{ECO} < S_{BH}(A_E) - \frac{n}{2} \log S_{BH}(A_E) + \cdots, \label{holent}
\end{eqnarray}
where, $n$ is an integer; $n=3$ for non-rotating quantum isolated horizons. For rotating isolated horizons, there is some debate as to what precisely the correct coefficient of the logarithmic correction is, although there are arguments to the effect that it should be the same as non-rotating case.
The inequality (\ref{holent}) turns the inequality (\ref{g2l}) into 
\begin{eqnarray}
S_{bh}(A_E) + S_{GW} > S_{bh1}(A_1) + S_{bh2}(A_2)  . \label{g2l2}
\end{eqnarray}
This can be alternatively expressed as, using the quantum-corrected black hole entropy, as
\begin{eqnarray}
\frac{\exp{\bar A_E}}{{\bar A_E}^{n/2}} >  \frac{\exp({\bar A_1} + {\bar A_2} - S_{GW})}{({\bar A_1}{\bar A_2})^{n/2} }~, \label{entbd} 
\end{eqnarray} 
where, ${\bar A} \equiv S_{BH}(A) = A/4A_{Pl}$, with  $A_{Pl}$ being the Planck area . 

The entropy bound (\ref{entbd}) which holds for the compact remnant in binary black hole mergers, regardless of its actual astrophysical structure, can be solved to yield a minimal cross-sectional area of the compact merger remnant, given the measured horizon areas of the two inspiralling black holes and the entropy of the emitted gravitational waves. 

\subsection{Generalization to Exotic Compact Objects} 

Does a bound like that in (\ref{g2l2}) generalize to the situation where, not only the post-merger remnant but the inspiralling binary is also constituted by exotic compact objects which are not necessarily black holes ? We demonstrate that the answer is in the affirmative. Let us suppose that the inspiralling stars have entropies $S_{E1}$ and $S_{E2}$. If they merge into a compact object as considered above, the Generalized Second Law and the Holographic entropy bound together then predict that 
\begin{eqnarray}
S_{bh}(A_E) + S_{GW} > S_{E1} + S_{E2} ~.\label{entbdc}
\end{eqnarray}
On the other hand the Holographic bound by itself asserts that $S_{E1} < S_{bh}(A_{E1})~,~S_{E2} < S_{bh}(A_{E2})$, so that
\begin{eqnarray}
S_{E1} + S_{E2} < S_{bh}(A_{E1}) + S_{bh}(A_{E2})~. \label{hbd}
\end{eqnarray} 
The inequality (\ref{hbd}) is a {\it sufficient} condition for the validity of the bound (similar to  (\ref{g2l2})),
\begin{eqnarray}
S_{bh}(A_E) + S_{GW} > S_{bh}(A_{E1}) + S_{bh}(A_{E2}) ~. \label{fbd}
\end{eqnarray}
It may be considered as a {\it necessary} condition as well, from a physical standpoint, if one notices that its invalidation would rule out binary black hole merger to a black hole. 

We have argued therefore that the entropy bound (\ref{entbd}) holds, both for {\it all} possible compact binary mergers including but not restricted to black holes, as also for {\it all} possible compact post-merger remnants including but not restricted to black holes. Thus, the bound is {\it independent} of whether or not either the inspiralling binary system or the post-merger remnant consist of black holes.   

Since the gravitational wave entropy is a maximum for the equilibrium thermodynamic situation, we shall replace $S_{GW}$ in (\ref{entbd}) by its equilibrium value $S_{GW}^{EQ}$, so that the final formula reads,
\begin{eqnarray}
\frac{\exp{\bar A_E}}{{\bar A_E}^{n/2}} >  \frac{\exp({\bar A_{E1}} + {\bar A_{E2}} - S_{GW}^{EQ})}{({\bar A_{E1}}{\bar A_{E2}})^{n/2}}~, \label{entbd2} 
\end{eqnarray}
{\it Observe that the area lower bound (\ref{entbd2}) is of maximal generality for compact binary inspiral coalescing into a compact remnant. It is valid irrespective of whether these are black holes, neutron stars, white dwarfs or exotic compact objects like gravastars or boson stars.}  

If  $R_E$ characterizes the average linear size of such an object, and $r_{SE}$ its Schwarzschild radius, then one can define a {\it compactness ratio} $C_E \equiv R_E/r_{SE}$. This ratio is close to unity for a black hole, is about ${\cal O}(10)$ for a neutron star, ${\cal O}(10^3)$ for a white dwarf, and so on. It then follows that 
\begin{eqnarray}
A_E(R_E) = A(C_E r_{SE}) = C_E^2 A_E(r_{SE}) = C_E^2 M_E^2
\end{eqnarray}    
where $M_E$ is the mass of the compact object, and the last equality holds for very slowly-spinning compact objects. This enables us to reexpress the inequality (\ref{entbd2}) as a lower bound for the cross-sectional area of a compact remnant, expressed in terms of the scaled areas (functions only of the Schwarzschild radii) and compactness ratios of the inspiralling  binary as well as the post-merger remnant,
\begin{eqnarray}
\frac{\exp[{C_E^2 {\bar A}_E}]}{(C_E^2 {\bar A}_E)^{n/2}} &>& \frac{\exp [C_{E1}^2 {\bar A}_{E1} + C_{E2}^2 {\bar A}_{E2} - S_{GW}^{EQ}]}{( C_{E1}^2 {\bar A}_{E1} C_{E2}^2 {\bar A}_{E2})^{n/2}} \label{entbdg} 
\end{eqnarray}
In case the binary inspiral and the remnant are all spinning very slowly, a simplified inequality emerges 
\begin{eqnarray}
\frac{\exp[C_E^2 {\bar M}_E^2]}{(C_E {\bar M}_E)^n} >   \frac{\exp[C_{E1}^2 M_{E1}^2 + C_{E2}^2 M_{E2}^2 - S_{GW}^{EQ}]}{( C_{E1} M_{E1} C_{E2} M_{E2})^n}~. \label{entbdnr}
\end{eqnarray}
For the case of GW150914, the minimal area inequality is given by the {\it lhs} of (\ref{entbdnr}) and the {\it rhs} of (\ref{entbdg}), since the data points to a rather slowly spinning post-merger remnant. Another reason why this approximation may be applicable is that in our considerations, the `merger black box' is not at all relevant, since we deal with the coalescing binary when the components are far apart, and the remnant after stabilization long after the merger has taken place. There are no other sources of strong gravity in either of these situations to cause any appreciable tidal deformation (which has the potential to render this approximation invalid).  

\section{Estmating Gravitational Wave Entropy} 

To estimate the entropy of a gravitational wave signal as observed for instance at aLIGO for GW150914 or subsequent observations of binary compact mergers, we follow the statistical method proposed by Ma \cite{ma1985}. In this approach, the entropy of a dynamical system is determined from the {\it trajectory of motion}, i.e., certain general properties of the space of solution configurations of the system. This solution space is divided into groups $\Omega_{\lambda}$, where $\lambda$ is a real positive number. Each such group of the configuration space has a volume $\Gamma_{\lambda}$. $p_{\lambda}$ is defined as the probability of occurrence of a trajectory belonging to the group $\lambda$, given by the fraction of the total time the system spends in this group. Given a configuration belonging to the group, the probability of finding a {\it coincident} configuration in the group $\lambda$ is given by $\Gamma^{-1}_{\lambda}$. The {\it coincidence probability} of occurrence of two identical configurations in the same group is then given by the ratio $p_{\lambda} / \Gamma_{\lambda}$. The entropy is then defined \cite{ma1985} by the average  $S = \sum_{\lambda} p_{\lambda} \log (\Gamma_{\lambda}/p_{\lambda})$ over all groups $\lambda$.       

To transcribe this to the case of gravitational waves, we can label gravitational wave configuration groups by their Fourier modes characterized by the frequency $\omega$. We shall then define $p(\omega)$ as the probability of occurrence of a gravitational wave mode with frequency $\omega$. In terms of the intensities of the gravitational wave signal, we can identify $p(\omega)$ with the ratio $I(\omega) /I_0$, where, $I(\omega)$ is the spectral intensity of gravitational waves with frequency $\omega$, and $I_0$ is the total {\it integrated} intensity. Let $N_t(\omega)$ be the total number of gravitational wave modes with frequency $\omega$. One can then adapt the definition of entropy given in ref. \cite{ma1985} for gravitational waves
\begin{eqnarray}
S_{GW}^{EQ} = \int_{{\omega}_1}^{{\omega}_2} \frac{d\omega}{2\pi} p(\omega) \log \left[ \frac{\Gamma(\omega)}{p(\omega)} \right] ~\label{entr}
\end{eqnarray}
Since the spectral distribution is continuous, the transcription of `coincident configurations' is not easy, as it is in the discrete case. Instead, here we adopt the feature of {\it coherence} to enable that aspect. Thus, we consider gravitational wave mode pairs with the same frequency $\omega$ which are {\it coherent} in the sense of wave optics. If $N_c(\omega)$ is the number of such coherent gravitational wave mode pairs, the fraction $N_c(\omega) /N_t(\omega)$ must be the probability that, given a mode with the frequency $\omega$, there is another mode with the same frequency which is coherent with it. We can then identify $\Gamma^{-1}(\omega) \equiv N_c(\omega) / N_t(\omega)$, so that the gravitational wave analogue of the coincidence probability is the quantity $p(\omega)\Gamma^{-1}(\omega) = N_c(\omega)/N_0$. The number $N_0$ is proportional to the total integrated intensity. Eqn. (\ref{entr}) can then be rewritten
\begin{eqnarray}
S_{GW}^{EQ} = \int_{{\omega}_1}^{{\omega}_2} \frac{d\omega}{2\pi} \frac{I(\omega)}{I_0} \log \left[ \frac{N_0}{N_c(\omega)} \right]
\end{eqnarray} 

How do we estimate $N_c(\omega)$ ? As for electromagnetic fields, the entropy is maximised when the coherence is a minimum \cite{manwol65}. This implies that equilibrium ensues when $N_c(\omega) << N_t(\omega) << N_0$, i.e., $\Gamma(\omega)/p(\omega) >> 1~ \forall ~\omega \in [\omega_1,\omega_2]$. An estimate of $N_0$ is the ratio of the total energy radiated ${\cal E}_{GW}$ to  the energy $\hbar \omega$ of a graviton. If, as per the GW150914 datasheet, ${\cal E}$ is taken to be about $3~M_{\odot}~c^2 \simeq 9 \times 10^{53} ~ ergs$, and $\omega \approx 150 Hz$, then $N_0 \simeq 2.7 \times 10^{77}$. Since the entropy is maximised for the maximally incoherent gravitational wave signal, we can take $N_c(\omega) \sim 1~ \forall~ \omega$. With these approximations, an approximate estimate of the gravitational wave entropy can be derived. A more detailed numerical estimate, using Fourier transforms of the GW150914 signal, changes the result only marginally. 

This estimate of the maximum entropy of gravitational waves emitted during a compact binary coalescence is admittedly {\it heuristic} in absence of a more rigorous approach. One may argue that the gravitational wave entropy actually vanishes classically, where a two-body inspiral should lead to a monochromatic signal with vanishing entropy. Likewise,  quantum mechanically, where the gravitational wave emitted corresponds to a coherent state peaked around the classical solution, it is a pure state with vanishing entropy. However, this argument misses out on the possible complexity involved during the {\it merger process} of an inspiralling binary of compact stars, perhaps even black holes. So  long as the inspiralling compact stars are sufficiently far apart long before merger, indeed all one receives is a two-body monochromatic GW signal involving lower orders of a gravitational multipole expansion of slowly moving gravitating sources. Such a signal will indeed have a very low entropy. But that changes close to merger, when the inspiralling sources move rapidly, resulting in a violently dynamical spacetime with the emitted gravitational waves exhibiting extra features beyond the simplistic description in terms of either a two-body signal or a coherent quantum state. It is well known that in such a dynamical spacetime, particle creation must occur \cite{parker2009} with uncertainties in the particle number in the initial vacuum state. Thus, creation of gravitons is very likely towards the end-stage of merger, leading to entropy production. This is where the waves are not benignly coherent any more, and hence not a pure state quantum mechanically, and detailed analyses, beyond our considerations here, are warranted, for a more precise computation of the entropy carried by the emitted gravitational waves. The major part of the GW entropy arises from this very dynamical situation, from our standpoint. Since our interest is in the lower bound on the remnant area, we have taken the maximum possible entropy of the GW signal, treated as a wave with minimal coherence.       

\section{Results} 

In this section, we consider data from the observations of compact binary mergers from the LIGO-VIRGO collaboration, to demonstrate the scope and applicability of our main theoretical contentions. The approach in this section is more towards illustrative {\it proofs of principle} rather than elaborate analyses of the merger data, much of which is currently in the process of being garnered. The novelty in our approach is that one does not need data on tidal deformations of the coalescing neutron stars which is used in standard approaches to constrain NS EoS's. Instead, our lower bound on the remnant area is adequate to determine the validity of specific EoS's. 
 
\subsection{Holographic Area Bound for Binary Black Hole Merger}

The minimal area $A_E$ which saturates the bound (\ref{entbd2}) can be computed in terms of the {\it Lambert} $W$ function \cite{lamb96}. Numerical values of the $W$ function can be used for illustrative visualization of the minimal area profile as a function of masses, for binary black hole (BBH) mergers (where $C_{E1}=C_{E2}=1$), or binary neutron star mergers (BNS, $C_{E1}=C_{E2} =5$) to a remnant black hole ($C_E=1$), as exhibited in Fig.s 1 and 2 respectively. The estimate of the gravitational wave entropy is based on the LIGO GW150914 data sheet showing a signal with a peak frequency of 150 Hz. This leads to an estimated entropy of ${\cal O}(50)$ in units of the Boltzmann constant.  
\begin{figure}[h]
\begin{center}
\includegraphics[width=7cm]{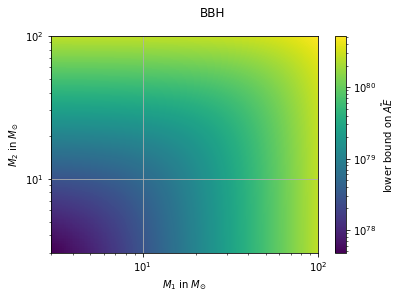}
\label{Fig.1}
\caption{Minimal remnant area vs coalescing masses for BBH merger}
\end{center}
\end{figure}
\begin{figure}[h]
\begin{center}
\includegraphics[width=7cm]{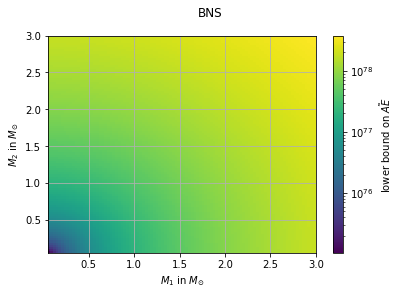}
 \label{Fig.2}
\caption {Minimal remnant area vs coalescing masses for BNS merger}
\end{center}
\end{figure}
\begin{figure}[h]
\begin{center}
 \includegraphics[width=7.5cm,height=5.5cm]{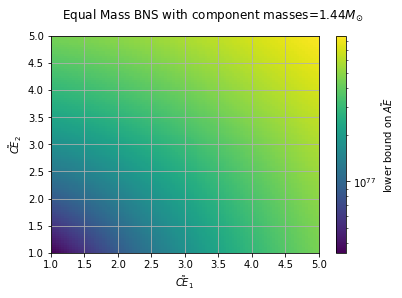}
 \label{Fig.3}
\caption { Compactness ratio $C_E$ vs $C_{E1}~,~C_{E2}$}
\end{center}
\end{figure}

Fig.s 1 and 2 show a similar {\it increasing} behaviour of the minimal remnant area as a function of the masses of the components in the binary merger. Thus, these plots verify our contention that irrespective of the astrophysical nature of the coalescing compact binary or the compact remnant, the minimal remnant area has a robust behaviour as a function of the  merging masses. For BBH merger to a black hole or even an ECO, accurate measurement of the radius or cross-sectional area remains a huge observational challenge, so at this stage of development, the plot in fig. 1 is a prediction for the future. If, for masses in the given ranges, a BBH merger produces a remnant with a {\it smaller} measured area than that predicted by the plot in fig. 1, then our estimated bound on minimal remnant area would be clearly invalidated. In other words, a more compact remnant has a greater likelihood of challenging our bound. 

For masses of the inspiral as well as the post-coalescence remnant close to the LIGO-VIRGO O2 data for BNS merger, one can plot the minimal value  of the compactness ratio $C_E$ of the remnant, as a function of those of the binary inspiral. This is shown in Fig.3.   

\subsection{Using the Holographic Area Bound to constrain Neutron Star Equations of State}

One important application of the holographic lower bound on compact binary merger remnants is the ability to constrain neutron star Equations of State, based solely on the observed masses of the components and the remnant. Unlike other constraints on Neutron star EoS based on only GW data (see for example \cite{LIGO6}) our constraint is independent of whether or not we can measure the individual tidal deformabilities of the components and the remnant. To compute this constrain, we can use the holographic bound in two different ways. 

Firstly, given the masses of the BNS merger and an EoS, we can compute the radii of the coalescing components and the remnant and hence the corresponding cross-sectional areas, under the spherical symmetry assumption $(A_{Ei}=4\pi r^{2}(m_i)$, say, for simplicity. If the remnant area corresponding to an observed remnant mass falls below the minimal remnant area obtained using these radii and eqn (\ref{entbd2}), then we would infer that the particular EoS is untenable. Thus, the component and remnant radii which have been obtained using EOS from the observed masses, expressed as $r_i=r_i(m_i|EOS)$, it must satisfy the inequality (all masses and radii in the formulae are in Planckian units in this subsection) :
\begin{eqnarray}
\frac{\exp{4\pi \tilde{r}^2}}{(4\pi \tilde{r}^2)^{n/2}} &>& \frac{\exp{\{4\pi \tilde{r}_1^2 + 4\pi \tilde{r}_2^2 - S_{GW}\}}}{\{4\pi \tilde{r}_1 \tilde{r}_2 \}^{n}} \label{minar}
\end{eqnarray}

We use the LIGO Arithmetic Library (LAL) Simulations \cite{lalsim} to compute the radii of merging neutron stars, given an EoS and observed masses, and construct a table exhibiting the radii of the components and the corresponding lower bound on the remnant radius. This table may serve as a proof of principle of our proposition: use the holographic bound and measurement of compact binary merger component and remnant masses to rule out EOS's. We choose our fiducial component masses to be close to that measurements of GW170817 \cite{LIGO5}. The results are plotted in Table 1 below. Observe that the table misses the column corresponding to the {\it measured} radius of the remnant, for any EoS. Had the remnant mass of GW170817 been explicitly measurable, the table would have sufficed to rule out EOS's that yield an observed remnant radius {\it lower} than that predicted by our lower bound. While it is true that GW data on 170817 is insufficient to verify the remnant's existence, let alone constrain its mass \cite{LIGO5} , there are neutron star merger scenarios that predict a neutron star merger remnant, even if short lived (see for example \cite{metz},\cite{fab}). Table 1 below is ilustrative of our approach, and not exhaustive, the parameters have been computed for the entire list of neutron star EoS's, but not included here.
\vglue .2cm
\begin{center}
\begin{tabular}{|c|c|c|c|c|}
\hline
{\bf EoS} & {\bf r$_1$ (km)} & {\bf r$_1$ (km)} & {\bf Min r$_{rem}$ (km)} & {\bf Obs r$_{rem}$} \\
\hline
{\bf ALF1} & 9.09379 & 7.57538 &	12.86056 & \\
\hline 
{\bf AP3} & 12.10244 & 12.07771 &	17.11544 & \\
\hline
{\bf BSK20} & 11.73223 & 11.77605 &	16.59188 & \\
\hline
{\bf FPS}	& 8.33829 &	9.16387 & 11.79213 & \\
\hline
{\bf GNH3} & 14.36108 &	14.64615 &	20.30963 & \\
\hline
{\bf H5} &	13.05519 &	13.29034 &	18.46283 & \\
\hline
\end{tabular}
\end{center}

\noindent {\it TABLE 1. We compute lower bounds on the remnant radius given the component radii. LAL Simulations Neutron Star Mass-Radius Functions are used to get the radii of the components given their masses for each EoS. The component masses used Results for six randomly selected EoS's have been entered for illustrative purposes. Observed remnant radii are yet unavailable from GW170817 data. }
\vglue .2cm

Another approach involves the masses and an EoS, from which we can compute the dimensionless compactness rato ($C_E$) of the components and the remnant. This can be substituted into the inequality eqn (\ref{entbdg}); the EoS's for which (\ref{entbdg}) is not satisfied, may be ruled out. For an EoS to satisfy our constraint based on our holographic bound, where the radius has been obtained using it from the mass, expressed as $r_i=r_i(m_i)$, it must satisfy the inequality :
\begin{eqnarray}    
\frac{\exp{\{C^2_E\tilde{M}_E^2 \}}}{(C_E \tilde{M}_E)^n} &>& \frac{\exp{\{C^2_{E1} \tilde{M}_{E1}^2+C^2_{E2} \tilde{M}_{E2}^2 - S_{GW}\}}} {[C_{E1} \tilde{M}_{E1}C_{E1}\tilde{M}_{E1}]^n} \label{comp}
\end{eqnarray}
In case the merging binary components are slowly-spinning, the results of the two approaches overlap, giving a lower bound on the remnant properties that is independent of the measurement of tidal deformabilities of the remnants and components. A measurement of the component masses (which is extremely viable as per \cite{LIGO5}) and the remnant mass is enough to rule out EOS's based on eqns (1) and (2) (in this tex file). Hence future detection of a BNS event with a remnant of measurable mass remains a strong possibility. We conclude that given such a possibility, our frame work provides an elegant method to constrain the EoS of matter in neutron stars based on only mass measurements and theoretical considerations.

\section{Concluding Remarks} 

The possibility of contraining neutron star EoS's on the basis of the lower bound on the remnant area, without reference to data on tidal deformations of the coalescing components, is an interesting development on its own right. While at this stage only a proof of principle could be provided, future, more accurate observations on remnant masses in BNS mergers may provide the opportunity for detailed numerical analyses towards ruling out proposed neutron star EoS's of neutron star matter.  

One of the key issues is the significance of the LQG quantum corrections incorporated into our formulae eqn.s (\ref{entbd}) - (\ref{entbd2}). These corrections have become related to measureable quantities through our illustrative plots and tables. These corrections may not have a lot of significance at the current precision of measurements in gravitational wave interference experiments, but as observational accuracy increases with time, it is conceiveable that their role can be subject to measurement. 

We have already remarked on the allegedly `heuristic' nature of the Bekenstein entropy bound, quintessentially used in this paper. In a later paper \cite{bekbd81} following his incipient 1974 proposal \cite{bekbd74} of the entropy bound, Bekenstein had proposed a {\it universal} bound on the entropy of all bounded objects in flat spacetime, once again based on thermodynamic considerations. In 2008, Casini \cite{cas08} has formulated the notion of {\it relative entropy}, whose positivity has been shown to lead to a more rigorous derivation of the bound in flat spacetime. A generalization of such ideas to Loop Quantum Gravity would be a key project for the future.

\noindent{\it Acknowledgment :} We thank an anonymous referee for  bringing ref.\cite{lamb96} to our attention. One of us (PM) thanks E. Witten for bringing to his attention ref. \cite{cas08}, in one of his recent online lectures and in correspondence following it.


\end{document}